\documentclass[12pt]{article}

\usepackage{amsmath, amssymb}
\usepackage{subeqnarray}
\newcommand{\hin}{\rangle\!}
\newcommand{\vor}{\!\langle}
\newcommand{{\be}}{{\begin{equation}}}
\newcommand{{\ee}}{{\end{equation}}}

\begin{document}

\def\theequation{\thesection.\arabic{equation}}
\setcounter{equation}{0}

\title{Jordan Blocks and Exponentially
Decaying Higher Order Gamow States\thanks{Dedicated to L.~C. Horwitz
on the occasion of his 65$^{\rm th}$ birthday.}}
\author{{\bf A. Bohm}, {\bf M. Loewe}, {\bf P. Patuleanu}, 
and {\bf C. P\"{u}ntmann} \smallskip
\vspace{4pt}\\
{\footnotesize{Center for Particle Physics, 
Department of Physics}}\\ 
{\footnotesize{University of Texas at Austin, Austin, Texas 78712}}
}
\maketitle


\begin{abstract}

In the framework of the rigged Hilbert space, unstable quantum systems
associated with first order poles of the analytically continued
S-matrix can be described by Gamow vectors which are generalized vectors with
exponential decay and a Breit-Wigner energy distribution. 
This mathematical formalism can be generalized to quasistationary systems
associated with higher order poles of the S-matrix, which leads to a
set of Gamow vectors of higher order with a non-exponential time
evolution. One can define a state operator from the set of higher order
Gamow vectors which obeys the exponential decay law. We shall discuss
to what extend the requirement of an exponential time evolution
determines the form of the state operator for a quasistationary
microphysical system associated with a higher order pole of the S-matrix.
\end{abstract}




\section{Introduction}\label{introduction}\setcounter{equation}{0}

Gamow vectors in rigged Hilbert spaces (RHS) were introduced~\cite{BG} to
describe resonances, since they possess all the features usually
attributed to resonance states, in particular, they obey an
exponential decay law and have a
Breit-Wigner energy distribution. They can be constructed from the
first order poles of the analytically continued S-matrix on the
second Riemann sheet of the complex energy plane, and are 
generalized eigenvectors of a self-adjoint
Hamiltonian with complex eigenvalues (energy and lifetime).
Considering the S-matrix poles on the second Riemann sheet 
of higher order, e.g. order $r$, one can construct in the same way $r$ 
Gamow vectors of orders $k=0,~1,~2,\dots,r-1$.
These higher order Gamow vectors are Jordan vectors of degree
$k+1$,~\cite{antoniou-gadella} and they span an $r$-dimensional
subspace of ${\cal M}_{z_R}\subset\Phi^\times$, where $\Phi^\times$ is
the space of generalized vectors (functionals) of the Rigged Hilbert space
$\Phi\subset{\cal H}\subset\Phi^\times$. For the self-adjoint
Hamiltonian $H$ one obtains in the space $\Phi^\times$ a matrix
representation which, when restricted to the subspace ${\cal M}_{z_R}$,
form a Jordan block of degree r.~\cite{BKGL} 

Jordan block matrices for the Hamiltonian were already investigated by
Katznelson~\cite{katznelson} who based his work on the well
established ideas of Horwitz ~\cite{horwitz-marchand}. However, their
Hamiltonians are 
non-self-adjoint, and their matrices are a generalization of the
standard finite dimensional complex diagonalizable matrices for
``effective'' non-hermitian Hamiltonians~\cite{LWY}. This
approach cannot be implemented in a quantum 
theoretical framework using the Hilbert space. In contrast, the
Gamow-Jordan vectors that are derived from higher order S-matrix poles
have, like ordinary Gamow vectors representing Breit-Wigner resonances,
a natural place in the rigged Hilbert space formulation of quantum mechanics.

In section~\ref{sec:resonances}, we will give a brief introduction to
rigged Hilbert spaces, ordinary and higher order Gamow vectors, 
and their time evolution. In section~\ref{sec:uniqueness}, we will 
define the exponentially decaying higher order Gamow state 
operator which was conjectured in reference~\cite{ourselves} and
discuss its generalization.
We argue that these higher order Gamow state operators
are of this form because (with certain qualifications) this is the
only way to reconcile them with the exponential law.


\section{Resonances, Rigged Hilbert Spaces, \\ and Gamow Vectors}
\label{sec:resonances}\setcounter{equation}{0}

Conventionally, a resonance is
described by a pair of first order poles of an analytically continued
S-matrix~\cite{BG}. These poles lie on the second sheet of the
two-sheeted Riemann surface of complex energy at the
complex conjugate positions $z_R=E_R-i\Gamma/2$ and
$z_R^*=E_R+i\Gamma/2$, where $E_R$ is the 
resonance energy and $\hbar/{\Gamma}$ is the lifetime of the
resonance. The pole in the lower half-plane can be associated with a
state that decays exponentially (for $t>0$), while the pole  
in the upper half-plane can be associated with a state that grows
exponentially (for $t<0$).
While in Hilbert space there are no such states that have an
exponential time evolution and distinguish a
direction of time, in the rigged Hilbert space (RHS) one describes
such states by antilinear continuous functionals, the Gamow vectors.

The rigged Hilbert space consists of a triplet of spaces~\cite{gelfand}:
\begin{equation}
	\Phi\subset{\cal H}\subset\Phi^\times\,.
\end{equation}
The best known example is the RHS where
$\Phi$ is the space (Schwartz space) of ``well-behaved'' functions,
i.e. functions, that have derivatives, which are all continuous, 
smooth, and rapidly decreasing.
${\cal H}$ is the space of Lebesgue square
integrable functions, and $\Phi^\times$ (the space of continous
antilinear functionals on $\Phi$) is  the space of tempered distributions.

In quantum theory, if one distinguishes between preparations and
registrations~\cite{ludwig-foundations}, one can further specify the
RHS, and is led to a pair of RHS's: one for the preparations and one 
for the registrations~\cite{cologne}.
A scattering experiment can be subdivided into a preparation stage and
a registration stage. The in-state $\phi^+$
that evolves from the prepared in-state $\phi^{\rm in}$ outside 
the interaction region is determined by the preparation
apparatus (the accelerator). The out-state $\psi^-$, detected as the
``out-state''  $\psi^{\rm out}$ outside the
interaction region, is determined by the registration apparatus (detector).
According to the physical interpretation of the RHS formulation,
``real'' physical entities connected with the experimental apparatuses,
e.g. the ensemble $|\phi\rangle\langle\phi|$ describing the
preparation apparatus (the energy distribution of the beam) or the 
observable $|\psi\rangle\langle\psi|$ describing the registration 
apparatus (the energy resolution of the detector) are described by 
the ``well-behaved'' vectors $\phi,~\psi\in\Phi$. One denotes the
space of state vectors $\phi^+$ by $\Phi_-$ and the space of
``observable vectors'' $\psi^-$ by $\Phi_+$, where $\Phi=\Phi_-+\Phi_+$
and $\Phi_-\cap\Phi_+\neq 0$. $\Phi_-$ is the space of
``well-behaved'' Hardy class vectors from below and $\Phi_+$ is the space of ``well-behaved'' Hardy class vectors from
above, i.e. they fulfill even stronger properties connected with their
analytic continuation, than the  elements of the Schwartz space~\cite{duren-hoffman}.
We will call these elements ``{\it very} well-behaved'' vectors. In
place of the single rigged Hilbert space we therefore have a pair of
rigged Hilbert spaces: 
\begin{eqnarray}
	\phi^+\in&\Phi_-\subset{\cal H}\subset\Phi_-^\times
	&\text{~~~for~ensembles~or~prepared~in-states,}\\
	\psi^-\in&\Phi_+\subset{\cal H}\subset\Phi_+^\times
	&\text{~~~for~observables~or~registered~``out-states''.}
\end{eqnarray}
Here the Hilbert space ${\cal H}$ is the same for both triplets. 

On the level of $\Phi$ or ${\cal H}$ one cannot talk of single 
microsystems, and there are no mathematical objects in
Hilbert space quantum mechanics to describe a single microsystem or a
single experiment which prepares and observes a single
microsystem. Still, it is intuitively attractive to imagine that the
effect by which the preparation apparatus acts on the registration
apparatus is carried out by single physical entities, the
microphysical systems. The energy
distribution for a microphysical system does not have to be a ``well-behaved'' 
function of the physical values of energy $E$. Hence, for the 
hypothetical entities connected with microphysical systems, like 
Dirac's ``scattering states'' $|{\rm {\bf p}}\rangle$ or Gamow's 
``decaying states'' $|E-i\Gamma/2\rangle$, the RHS formulation uses 
elements of $\Phi^\times$, $\Phi^\times_+$, and
$\Phi^\times_-$. 

The decaying Gamow vector associated with the complex energy 
$z_R=E_R-i\Gamma/2$ is a continuous antilinear functional 
over the vectors $\psi^-\in\Phi_+$, and its time evolution is defined for times $t\geq0$. They are generalized eigenvectors of (the extension of) a
self-adjoint Hamiltonian $H$ with complex eigenvalue $z_R$
\begin{equation}
	H^\times|z^-_R\rangle=z_R|z^-_R\rangle\,;\hspace{1cm}z_R=E_R-i\Gamma/2
\end{equation}
(where the $\,^\times$ denotes the conjugate operator acting on the 
functionals). These Gamow vectors can be obtained from the first order
poles of the  
analytically continued S-matrix at the position $z_R$ on the lower 
half-plane of the second Riemann sheet,
\begin{equation}
	S(z)=\frac{a_{-1}}{z-z_R} + \rm{~~~analytic~terms}\,.
\end{equation}
Here $a_{-1}$ can determined from the unitarity of the S-matrix to
be $-i\Gamma$~\cite{abohm}. In the same way one can define the
exponentially growing Gamow vector associated with the complex energy
$z^*_R=E_R+i\Gamma/2\,$, which is a continuous antilinear functionals
over the vectors $\phi^+\in\Phi_-$.

The $n^{\rm th}$ order Gamow vector associated with the complex energy
$z_R$ is a generalization of the Gamow vector above
in the following way:

\vspace{12pt}
1) There are $r$ generalized vectors (functionals) of order
$n=0,~1,~\dots,~r-1$. They are associated
with a pole of the order $r$ at the
position $z_R$ on the second Riemann sheet of the analytically
continued S-matrix,
\begin{equation}\label{s}
	S(z)=\frac{a_{-1}}{z-z_R}+\frac{a_{-2}}{(z-z_R)^2}+\dots+
 	\frac{a_{-r}}{(z-z_R)^{r}}+\rm{analytic~terms}
\end{equation}
where the coefficients $a_{-n-1},~n=0,~1,\dots,~r-1$ 
can be determined from the unitarity
of the S-matrix~\cite{ourselves} and are
complex numbers with dimension [energy]$^{n+1}$. 

One starts from the S-matrix elements,
i.e. the matrix elements of the incoming prepared
state $\phi^{\rm in}\in\Phi_-$ at $t\rightarrow-\infty$ and the
outgoing detected observable $\psi^{\rm out}\in\Phi_+$ at
$t\rightarrow\infty$
\begin{eqnarray}\nonumber
	(\psi^{\rm out},S\phi^{\rm in})
	&=&(\psi^{\rm out},{\Omega^-}^\dagger\Omega^+\phi^{\rm in})
	=(\Omega^-\psi^{\rm out},\Omega^+\phi^{\rm in})\\
	&=&(\psi^-,\phi^+)=\int_0^\infty\;
	dE\langle\psi^-|E^-\hin \,S(E)\,\vor^+E|\phi^+\rangle\;,
\end{eqnarray}
where $\Omega^\pm$ are the M{\o}ller wave operators.
One can now deform the contour of integration from the cut (positive
real axis and spectrum of $H$), into  the lower half-plane of the
second sheet. Then one obtains a ``background integral'' term, independent of the
poles, along the negative real axis of the second sheet and a residue
term for the pole: 
\begin{equation}
	(\psi^-,\phi^+)=\int_{-\infty_{II}}^0\;dE\langle\psi^-|E^-\hin
	\,S(E)\,\vor^+E|\phi^+\rangle+{\rm residue~at~} z_R.
\end{equation}
For every residue term of the S-matrix pole at $z_R$ one obtains:
\begin{eqnarray}\nonumber
	{\rm residue~at~} z_R&=&\sum_{n=0}^{r-1}
	\frac{-2\pi ia_{-n-1}}{n!}\frac{d^{n}}{dz^{n}}
	\left(\langle\psi^-|z^-\rangle\langle
	z^+|\phi^+\rangle\right) 
\\	&=&\sum_{n=0}^{r-1}\frac{-2\pi ia_{-n-1}}{n!}\sum_{k=0}^{n}\label{PPP}
	\left(\!\!\begin{array}{c}n\\k\end{array}\!\!\right)
	\left.\frac{d^{n-k}}{dz^{n-k}}\langle\psi^-|z^-\rangle
	\;\frac{d^{k}}{dz^{k}}\langle z^+|\phi^+\rangle\right|_{z=z_R}
\\	&=&\sum_{n=0}^{r-1}\frac{-2\pi ia_{-n-1}}{n!}\sum_{k=0}^{n}
	\left(\!\!\begin{array}{c}n\\k\end{array}\!\!\right)
	\langle\psi^-|z^-_R\hin^{(n-k)}\;^{(k)}\vor^+z_R|\phi^+\rangle
	\nonumber
\end{eqnarray}
where $\langle\psi^-|z^-_R\hin^{(n-k)}$ is the 
$(n-k)$-th derivative of the ``well-behaved'' (continuous, analytic,
smooth, rapidly decreasing) function $\langle\psi^-|z^-\rangle\in{\cal
H}^2_-\cap{\cal S}$ at the position $z_R$, 
and is therefore analytic in the lower half-plane,
and $^{(k)}\vor^+z_R|\phi^+\rangle$ is the $k$-th
derivative of the ``well-behaved'' analytic function $\langle^+
z|\phi^+\rangle\in{\cal H}^2_-\cap{\cal S}$ at the position $z_R$
in the lower half-plane. 
The S-matrix~(\ref{s}) is thus associated with a set of
$r$ generalized vectors,
\begin{equation}\label{bohm38}
       |z^-_R\hin^{(0)},~|z^-_R\hin^{(1)},\dots ,~|z^-_R\hin^{(r-1)}
\end{equation}
where $|z^-_R\hin^{(0)}$ is the ordinary Gamow vector. (The same
generalization can be carried out for the growing higher 
order Gamow vectors with complex energy $z^*_R$ which we shall not
discuss here.)

\vspace{12pt}
2) It can be shown that the higher order Gamow vectors are generalized eigenvectors of the 
self-adjoint Hamiltonian $H$ in the following sense
\begin{eqnarray}
	\vor H\psi^-|z_R^-\rangle&=\,\langle\psi^-|H^\times
	|z^-_R\hin^{(0)} \,=& z_R\,
	\langle\psi^-|z^-_R\hin^{(0)} \nonumber \\ 
	\langle H\psi^-|z_R^-\hin^{(1)}&=\,\langle\psi^-|H^\times
	|z^-_R\hin^{(1)} \,=& z_R\, \langle\psi^-|z^-_R\hin^{(1)} +\, 
	\langle\psi^-|z^-_R\hin^{(0)} \nonumber \\
	&\vdots&  \label{eq:48}\label{haha}\\
	\langle H\psi^-|z_R^-\hin^{(k)}&=\,\langle\psi^-|H^\times
	|z^-_R\hin^{(k)} \,=& z_R\, \langle\psi^-|z^-_R\hin^{(k)} +k\,
	\langle\psi^-|z^-_R\hin^{(k-1)} \nonumber \\
	&\vdots&  \nonumber \\
	\langle H\psi^-|z_R^-\hin^{(r-1)}&=\,\langle\psi^-|H^\times
	|z^-_R\hin^{(r-1)} \,=& z_R\,
	\langle\psi^-|z^-_R\hin^{(r-1)} +(r-1)\, 
	\langle\psi^-|z^-_R \hin^{(r-2)}\nonumber 
\end{eqnarray}
Omitting the arbitrary $\psi^-$ one writes this as
\begin{equation}
	H^\times|z_R^-\hin^{(k)}=z_R|z_R^-\hin^{(k)}+
	k|z_R^-\hin^{(k-1)}\,.
\end{equation}
From this it follows that $H^\times$ is a Jordan operator of degree $r$, and
the $k$-th order Gamow vector $|z^-_R\hin^{(k)}$ is a Jordan vector
of degree $k+1$~~\cite{BKGL}, i.e.,
\begin{equation}
	(H^\times-z_R)^{(k+1)}|z^-_R\hin^{(k)}=0 \,\,{\rm ~~~and~~~}\,\,
	(H^\times-z_R)^{(k+1)}|z^-_R\hin^{(k-1)}\neq0\;.
\end{equation}

3)  The time evolution of the decaying $k$-th order Gamow vector is given by
[2]
\begin{equation}
  	e^{-iH^\times t}|z_R^-\hin^{(k)}=e^{-iz_Rt}\sum_{p=0}^k
  	\left(\!\!\begin{array}{c}k\\p\end{array}\!\!\right)
  	\left(-it\right)^{k-p}|z_R^-\hin^{(p)},\hspace{10mm}
  	t\geq 0.
\end{equation}
The $k$-th order Gamow vector evolves into a superposition containing Gamow
vectors of the same and all lower orders and the space spanned by the set
(2.10) is thus invariant under the action of the time evolution operator.  It
follows from (2.14) that the time evolution of the dyadic product of a $k$-th
order Gamow vector with an $m$-th order Gamow vector, $|z_R^-
\hin^{(k)}\;^{(m)}\vor^-z_R|$, 
has terms with additional powers of $t$ multiplying the overall
exponential factor $e^{-\Gamma t}$.  Such non-exponential time
evolution is really no surprise in view of the previous results [14].  A more
surprising fact is that certain linear combinations of the dyadic products,
e.g.,
\begin{equation}\label{con0}
  	W^{(n)}\equiv{\Gamma^n\over n!}\sum_{k=0}^n
	\left(\!\!\begin{array}{c}n\\k\end{array}\!\!\right)
  	|z_R^-\hin^{(k)}\;^{(n-k)}\vor z_R|,\hspace{10mm} 
  	\hbox{for fixed $n\in\{0,1,\cdots,r-1\}$},
\end{equation}
have a purely exponential time evolution [7]:
\begin{equation}
  	W^{(n)}(t)=e^{-iH^\times t}W^{(n)}(0)e^{iHt}=e^{-\Gamma t}
  	W^{(n)}(0),\hspace{10mm}\hbox{for $t\geq 0$.}
\end{equation}

\section{Form of the Exponentially Decaying
Operators}\label{sec:uniqueness}
\setcounter{equation}{0}

We now wish to determine to what extent the requirement of exponential time
evolution restricts the form of an operator constructed as a linear combination
of dyadic products of vectors in ${\cal M}_{z_R}$.  The most general
linear combination of dyadic products of vectors in
${\cal M}_{z_R}$ is given by
\begin{equation}
  W
  =
  \sum_{k=0}^{r-1}
  \sum_{m=0}^{r-1}
  B_{m,k}
  |z_R^-\hin^{(k)}\;^{(m)}\vor^-z_R|
\end{equation}
with arbitrary coefficients $B_{m,k}$.  We will show that $W$ decays according
to the pure exponential $e^{-\Gamma t}$ if and only if the coefficients are
restricted by
\begin{equation}
	B_{m,k}=\left\{\begin{array}{cl}
		{\footnotesize{\left(\!\!\begin{array}{c}
		k+m\\k\end{array}\!\!\right)}}
		B_{k+m,0},& {\rm for}~k+m\leq r-1,\\
		0, & {\rm for}~k+m> r-1,
	\end{array}\right.
\end{equation}
where the coefficients $B_{k+m,0}$ remain 
arbitrary, that is, if and only if $W$ is restricted to the form
\begin{equation}
	W=\sum_{n=0}^{r-1}B_{n,0}\sum_{k=0}^n
  	\left(\!\!\begin{array}{c}n\\k\end{array}\!\!\right)
  	|z_R^-\hin^{(k)}\;^{(n-k)}\vor^-\!z_R|
\end{equation}
with arbitrary coefficients $B_{n,0}$.  Since the vectors
$|z_R^-\hin^{(k)}$ have the dimension $[{\rm
energy}]^{-{1\over2}-k}$, the sums over $k$ in (3.3) have the dimensions
$[{\rm energy}]^{-1-n}$ so that, if $2\pi\Gamma W$ is to be
dimensionless, then the coefficients $B_{n,0}$ must have the dimensions $[{\rm
energy}]^n$ as in (2.15).

For the proof it is convenient to consider the most general linear combination
of dyadic products $|z_R^-\hin^{(k)}\;^{(m)}\vor^-z_R|$ 
for which the sum $n=k+m$ of
the orders does not exceed a finite integer $j$; this is given by
\begin{equation}
  	W_{\!(j)}=\sum_{n=0}^j\sum_{k=0}^nA_{n,k}
	|z_R^-\hin^{(k)}\;^{(n-k)}\vor^-z_R|
\end{equation}
with arbitrary coefficients $A_{n,k}$.  The operator (3.1) is obtained
as a special case of the operator (3.4) by setting
\begin{equation}
  	j=2(r-1),\hspace{1cm}A_{n,k}=B_{n-k,k}= 
	\left\{\begin{array}{rc}
		0, {\rm ~~for} & n-k>r-1, \\
        	0, {\rm ~~for} & \hspace{6.2mm} k>r-1.
	\end{array}\right.
\end{equation}
The time dependence of $W_{(j)}$ is given, using (2.14), by
\begin{eqnarray}\nonumber
  	W_{(j)}(t)&=&\left(e^{iHt}\right)^{\times}W_{(j)}(0)e^{iHt}=
  	\sum_{n=0}^j\sum_{k=0}^nA_{n,k}\left(e^{iHt}\right)^{\times}
  	|z_R^-\hin^{(k)}\;^{(n-k)}\vor^-z_R|e^{iHt}\\
  	&=&e^{-\Gamma t}\sum_{n=0}^j\sum_{k=0}^n\sum_{l=0}^k\sum_{m=0}^{n-k}
  	A_{n,k}\left(\!\!\begin{array}{c}k\\l\end{array}\!\!\right)
	{n\!-\!k\choose m}\left(-it\right)^{k-l}\left(it\right)^{n-k-m}
  	|z_R^-\hin^{(l)}\;^{(m)}\vor^-\!z_R|.\nonumber
\end{eqnarray}
Changing the order of the summations,
\begin{equation}\nonumber
  \sum_{n=0}^j
  \sum_{k=0}^n
  \sum_{l=0}^k
  \sum_{m=0}^{n-k}
  \!\!=\!\!
  \sum_{n=0}^j
  \sum_{l=0}^n
  \sum_{k=l}^n
  \sum_{m=0}^{n-k}
  \!\!=\!\!
  \sum_{l=0}^j
  \sum_{n=l}^j
  \sum_{k=l}^n
  \sum_{m=0}^{n-k}
  \!\!=\!\!
  \sum_{l=0}^j
  \sum_{n=l}^j
  \sum_{m=0}^{n-l}
  \sum_{k=l}^{n-m}
  \!\!=\!\!
  \sum_{l=0}^j
  \sum_{m=0}^{j-l}
  \sum_{n=l+m}^j
  \sum_{k=l}^{n-m},
\end{equation}
allows the dyadic products, which are linearly independent operators, to be
factored out of the sums over terms in which they appear as common factors:
\begin{eqnarray}\nonumber
  W_{(j)}(t)
  &=&
  e^{-\Gamma t}
  \sum_{l=0}^j
  \sum_{m=0}^{j-l}
  \sum_{n=l+m}^j
  \sum_{k=l}^{n-m}
  A_{n,k}
  {k\choose l}
  {n\!-\!k\choose m}
  \left(-it\right)^{k-l}
  \left(it\right)^{n-k-m}
  |z_R^-\hin^{(l)}\;^{(m)}\vor^-z_R|
  \\\nonumber
  &=&
  e^{-\Gamma t}
  \sum_{l=0}^j
  \sum_{m=0}^{j-l}
  |z_R^-\hin^{(l)}\;^{(m)}\vor^-z_R|
  \sum_{n=l+m}^j
  \sum_{k=l}^{n-m}
  A_{n,k}
  {k\choose l}
  {n\!-\!k\choose m}
  \left(-it\right)^{k-l}
  \left(it\right)^{n-k-m}
  \\\nonumber
  &=&
  e^{-\Gamma t}
  \sum_{l=0}^j
  \sum_{m=0}^{j-l}
  |z_R^-\hin^{(l)}\;^{(m)}\vor^-z_R|
  \sum_{n=l+m}^j
  \left(it\right)^{n-m-l}
  \sum_{k=l}^{n-m}
  A_{n,k}
  {k\choose l}
  {n\!-\!k\choose m}
  \left(-1\right)^{k-l}.
\end{eqnarray}

The operator $W_{(j)}(t)$ will decay according to the pure exponential
$e^{-\Gamma t}$ if and only if all terms involving additional powers of $t$
cancel.  All terms involving additional powers of $t$ will cancel if and only
if the coefficients $A_{n,k}$ satisfy the conditions
\begin{equation}
  0
  =
  \sum_{k=l}^{n-m}
  A_{n,k}
  {k\choose l}
  {n\!-\!k\choose m}
  \left(-1\right)^{k-l}
  \hspace{10mm}\hbox{for}\hspace{10mm}
  \left\{\begin{array}{l}
  	l\in\{0,\cdots,j-1\}, \\
  	m\in\{0,\cdots,j-1-l\}, \\
  	n\in\{m+l+1,\cdots,j\},
  \end{array}\right.
\end{equation}
The simplest of these conditions are those for which $n=m+l+1$, i.e., those for
which $m=n-l-1$, because they are the only conditions that involve sums over
only two values of $k$:
\begin{eqnarray}\nonumber
  0
  &=&
  \sum_{k=l}^{l+1}
  A_{n,k}
  {k\choose l}
  {n\!-\!k\choose n\!-\!l\!-\!1}
  \left(-1\right)^{k-l}
  \\\nonumber
  &=&
  A_{n,l}
  {n\!-\!l\choose n\!-\!l\!-\!1}
  -
  A_{n,l+1}
  {l\!+\!1\choose l}
  \hspace{10mm}\hbox{for}\hspace{10mm}
  \left\{\begin{array}{l}
  	n\in\{1,\cdots,j\}, \\
  	l\in\{0,\cdots,n-1\} 
  \end{array}\right.
\end{eqnarray}
or, replacing $l$ with $k-1$,
\begin{equation}\nonumber
  A_{n,k-1}
  {n\!-\!k\!+\!1\choose n\!-\!k}
  =
  A_{n,k}
  {k\choose k\!-\!1}
  \hspace{10mm}\hbox{for}\hspace{10mm}
  \left\{\begin{array}{l}
  	n\in\{1,\cdots,j\}, \\
  	k\in\{1,\cdots,n\} 
  \end{array}\right.
\end{equation}
or, using the definition ${n\choose k} \equiv {n!\over k!(n-k)!}$,
\begin{equation}\nonumber
  A_{n,k}
  =
  {(n\!-\!k\!+\!1)!(k\!-\!1)!\over (n\!-\!k)!k!}
  A_{n,k-1}
  \hspace{10mm}\hbox{for}\hspace{10mm}
  \left\{\begin{array}{l}
  	n\in\{1,\cdots,j\}, \\
  	k\in\{1,\cdots,n\}. 
  \end{array}\right.
\end{equation}
These conditions relate pairs of coefficients $A_{n,k}$ having the same values
of $n$ and successive values of $k$.  For fixed $n\in\{1,2,\cdots,j\}$, they may
be used recursively to show that $A_{n,k}$ must equal $A_{n,0}$ multiplied by
${n\choose k}$:
\begin{eqnarray}\nonumber
  A_{n,k}
  &=&
  \left[{(n\!-\!k\!+\!1)!(k\!-\!1)!\over (n\!-\!k)!k!}\right]
  \left[{(n\!-\!k\!+\!2)!(k\!-\!2)!\over (n\!-\!k\!+\!1)!(k\!-\!1)!}\right]
  \cdots
  \left[{(n\!-\!1)!1!\over (n\!-\!2)!2!}\right]
  \left[{n!0!\over (n\!-\!1)!1!}\right]
  A_{n,0}
  \\
  &=&
  {n!0!\over (n\!-\!k)!k!}
  A_{n,0}
  =
  {n\choose k}
  A_{n,0}
  \hspace{1cm}\hbox{for}\hspace{1cm}
  \left\{\begin{array}{l}
  	n\in\{1,\cdots,j\}, \\
  	k\in\{1,\cdots,n\}. 
  \end{array}\right.
\end{eqnarray}
Substituting this result into the full set of conditions (3.6), using the
identity
\begin{equation}\nonumber
  {n\choose k}
  {k\choose l}
  {n\!-\!k\choose m}
  =
  {n\choose m}
  {n\!-\!m\choose l}
  {n\!-\!m\!-\!l\choose k\!-\!l},
\end{equation}
and then using the binomial formula gives
\begin{eqnarray}\nonumber
  0
  &=&
  A_{n,0}
  \sum_{k=l}^{n-m}
  {n\choose k}
  {k\choose l}
  {n\!-\!k\choose m}
  \left(-1\right)^{k-l}
  \\\nonumber
  &=&
  A_{n,0}
  {n\choose m}
  {n\!-\!m\choose l}
  \sum_{k=l}^{n-m}
  {n\!-\!m\!-\!l\choose k\!-\!l}
  \left(-1\right)^{k-l}
  \\\nonumber
  &=&
  A_{n,0}
  {n\choose m}
  {n\!-\!m\choose l}
  \sum_{k-l=0}^{n-m-l}
  {n\!-\!m\!-\!l\choose k\!-\!l}
  1^{n-m-k}
  \left(-1\right)^{k-l}
  \\\nonumber
  &=&
  A_{n,0}
  {n\choose m}
  {n\!-\!m\choose l}
  (1-1)^n
  \\\nonumber
  &=&
  A_{n,0}
  {n\choose m}
  {n\!-\!m\choose l}
  0^n
  \hspace{10mm}\hbox{for}\hspace{10mm}
  \left\{\begin{array}{l}
  	l\in\{0,\cdots,j-1\}, \cr
  	m\in\{0,\cdots,j-1-l\}, \cr
  	n\in\{m+l+1,\cdots,j\}, 
  \end{array}\right.
\end{eqnarray}
which shows that the remaining conditions are automatically satisfied by (3.7)
without placing any further conditions on the coefficients $A_{n,0}$.  The
coefficients $A_{n,0}$, for $n\in\{1,\cdots,j\}$, and also the coefficient
$A_{0,0}$, remain completely arbitrary.

We conclude that a linear combination of dyadic products 
$|z_R^-\hin^{(k)}\;^{(m)}\vor^-z_R|$ for which the sum $n=k+m$ of the
orders does not exceed $j$ decays according to the pure exponential
$e^{-\Gamma t}$ if and only if it is of the form
\begin{equation}\label{con2}
  \sum_{n=0}^j
  A_{n,0}
  \sum_{k=0}^n
  {n\choose k}
  |z_R^-\hin^{(k)}\;^{(n-k)}\vor z_R|
\end{equation}
with arbitrary coefficients $A_{n,0}$.  Equation (3.3), i.e., (3.1) with
the restrictions (3.2), follows by applying the special case (3.5) to the
result (3.8).  This concludes the proof.

\section{Summary and Conclusion}\label{sec:conclusions}\setcounter{equation}{0}

In the conventional Hilbert space quantum mechanics, resonance states
cannot be described by state vectors. Therefore the most common
definition of a resonance is as a pole of the S-matrix at the complex
energy $z_R=E_R-i\Gamma/2$ (actually, a pair of poles at $E_R\pm
i\Gamma/2$). There is no theoretical reason to exclude poles of
order higher than one, and poles of any order will lead to the typical
resonance phenomena for cross-section and phase shift, in particular,
the time delay and therewith formation of a quasi-stationary
state~\cite{abohm}. But higher order poles have been scorned, because they were
somehow associated with a time evolution that in addition to the
exponential had also a strong (of order $\hbar/\Gamma$) polynomial time
dependence~\cite{goldberger} for which there exists no experimental evidence.
In the rigged Hilbert space formulation of quantum mechanics, a vector
description of resonances is possible, and Gamow vectors and Gamow
states were defined from the first order pole of the S-matrix
element. Their Hamiltonian time evolution was shown to be exactly
exponential and given by a semigroup~\cite{bohmjmp}. This procedure was
generalized to an $r$-th order pole at $z=z_R$ which led to the
definition of $r$ Gamow vectors of order
$k=0,~1,\dots,~r-1$.~\cite{antoniou-gadella} ~ The Gamow vectors of
order $k\geq 1$ were shown to have indeed a polynomial time evolution
in addition to the exponential~\cite{antoniou-gadella}. However, one
can find a non-reducible state operator in the $r$-dimensional space
${\cal M}_{z_R}$ spanned by the $r$ Gamow vectors, e.g. the
operator~(\ref{con0}), which has a purely exponential time
evolution~\cite{ourselves}. In the paper at hand 
we investigated the question, to what extend the requirement of a 
purely exponential time evolution
with a lifetime $\tau=\hbar/\Gamma=\hbar/{\rm Im}z_R$ determines the
form of the state opeerator. From rather plausible
assumptions we concluded that this must be a
``mixture'' of operators like~(\ref{con0}) with arbitrary
coefficients $A_{n,0}$ as given by~(\ref{con2}).


\bibliographystyle{ieeetr}

\end{document}